# Indicators of Open Access for universities

Nicolas Robinson-Garcia[1], Rodrigo Costas[2] and Thed N. van Leeuwen[3]

*[1] elrobinster@gmail.com*
Delft Institute of Applied Mathematics (DIAM), TU Delft, Netherlands

*[2] rcostas@cwts.leidenuniv.nl*
Centre for Science and Technology Studies (CWTS), Leiden University, Netherlands
DST-NRF Centre of Excellence in Scientometrics and Science, Technology and Innovation Policy, Stellenbosch University, South Africa

*[3] leeuwen@cwts.leidenuniv.nl*
Centre for Science and Technology Studies (CWTS), Leiden University, Netherlands

**Abstract**
This paper presents a first attempt to analyse Open Access integration at the institutional level. For this, we combine information from Unpaywall and the Leiden Ranking to offer basic OA indicators for universities. We calculate the overall number of Open Access publications for 930 universities worldwide. OA indicators are also disaggregated by green, gold and hybrid Open Access. We then explore differences between and within countries and offer a general ranking of universities based on the proportion of their output which is openly accessible.

**Introduction**

The recent announcement by Science Europe of Plan S, an initiative aiming at providing open access to all publications funded by a group of funding agencies (Else, 2018a, 2018b), has refuelled interest on Open Access at all levels. While Open Access (OA) has been on the agenda of the European Commission for quite some time now (Moedas, 2015), their favourable position towards implementing Plan S (Rabesandratana, 2019) invites to believe that it will soon be also mandatory for all EU funded research. The strictness of Plan S requirements, raises doubts on its viability (Frantsvåg & Strømme, 2019). But still, it evidences the need to monitor OA compliance at all levels, including institutional level.

Universities have been supporting OA for many years now. The most common has been by building and maintaining institutional repositories, and introducing mandates that oblige their researchers to deposit their publications (Harnad, 2007; Harnad et al., 2008). Another more recent way by which institutions are promoting OA, is by sponsoring costs derived from the article processing charges (APC) of open journals (Gorraiz & Wieland, 2009; Gorraiz, Wieland, & Gumpenberger, 2012). In any case, institutions are still faced with the challenge of determining how much of the research they produce is actually openly accessible. Initiatives such as the ranking of OA repositories (Aguillo, Ortega, Fernández, & Utrilla, 2010) offer a partial information which, although valuable, is still insufficient. One of the main limitations is that researchers may combine green and gold OA, and even when self-archiving their publications, they may deposit them in different repositories, impeding institutions to track efficiently their output.

Until recently, there were no more than estimates as to the amount of publications which were available in open access. But the development of platforms like CrossRef, DOAJ or even Google Scholar, along with computational advancements on web scraping, have led to a plethora of large-scale analyses to empirically identify OA literature (Archambault et al., 2014; van Leeuwen, Tatum, & Wouters, 2018; Martín-Martín, Costas, van Leeuwen, & Delgado López-Cózar, 2018; Piwowar et al., 2018). Overall, these studies report that around half of the scientific literature is freely available, but point towards the increasing availability of publications which do not adhere strictly to what is considered OA.

Here we highlight Unpaywall (Piwowar et al., 2018), which has had a great impact after being implemented by most of the main bibliometric data providers (Else, 2018c). Furthermore, the fact that the Unpaywall API can be freely queried allows others to assess on its performance but also to build on it. In this paper, we present a first attempt at analyzing Open Access at the institutional level, not only in general, but also focusing on the two main routes of OA: the green and the gold route; plus hybrid OA. The purpose for doing so is not only to inform university managers and funding agencies on the level of OA implementation of universities, but also to be able to understand and analyse national trends, and institutional strategies to implementing OA. Although we identify bronze OA, we exclude from our analyses due to the issues related with the sustainability of this type of OA, raising doubts as to its viability from a policy perspective (van Leeuwen, Meijer, Yegros-Yegros, & Costas, 2017). The results of this study have been recently incorporated to the 2019 edition of the Leiden Ranking released last May (van Leeuwen, Costas, & Robinson-Garcia, 2019).

**Data and methods**

In this paper we use different sets of sources and combine different methods to determine Open Access. The set of universities analysed and the identification of their publications is retrieved from CWTS in-house version of the Web of Science, based on the institutional name disambiguation developed to produce the Leiden Ranking (Waltman et al., 2012). For each publication, we identify if they are openly accessible and the type of Open Access by querying the Unpaywall information. The Unpaywall API does not labels types of OA but describes what information was derived from each record. More information on the information provided is available at their website the User Guide offered for researchers (http://unpaywall.org/data-format).

Four types of OA were considered. These four types of OA are defined as follows:

- **Green Open Access**. Self-archived versions of a manuscript. Here the responsibility lies on the author who is in charge of depositing the document in a repository. This version of the document may not correspond with the final version of the publisher.
- **Gold Open Access**. This refers to journals which publish all of their manuscripts in Open Access regardless of the business model they follow (e.g., publicly sponsored, author pays).
- **Bronze Open Access**. While again journals are the ones offering the publication freely available, this is not subjected to copyright conditions set to be defined as Open Access (i.e., they do not ensure perpetual free access).
- **Hybrid Open Access**. Non-OA journals make specific publications openly accessible usually after the author pays a fee to account for potential losses derived from subscription fees.

The labelling of OA types is described in Figure 1. It shows the workflow followed by evidence found for each publication, highlighting some of the difficulties and controversies raised when trying to define what is actually Open Access (Torres-Salinas, Robinson-Garcia, & Moed, 2019). Unpaywall API provides for each publication record a set of evidences of OA. For each evidence we would query the metadata labels as shown in Figure 1. When one evidence suggested that a paper belonged to an OA journal (gold OA), it automatically override further evidences of bronze or hybrid OA. The only exception made with green OA, which could overlap with any of the other three types.

**Figure 1. Workflow of OA labelling based on evidences by record found in Unpaywall. Percentages refer to OA share for publications from 963 universities in the 2014-2017 period.**

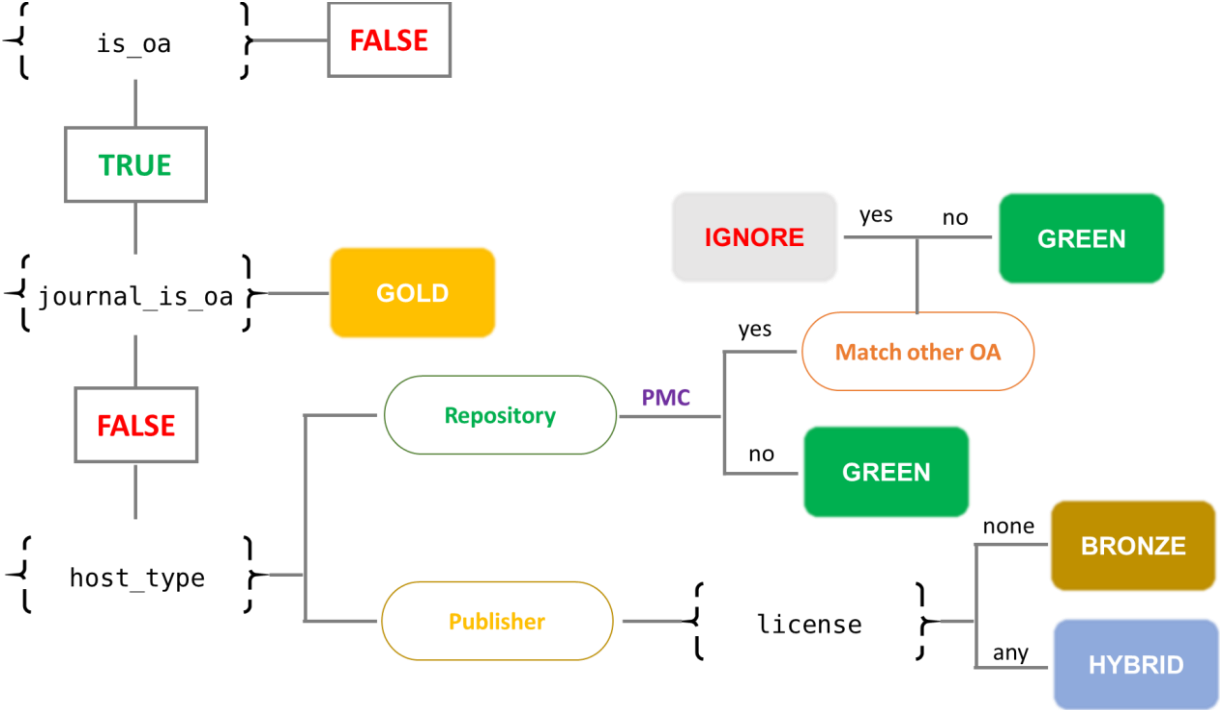

In all, a total of 963 universities were analysed for the 2014-2017 period. These are universities above the 1000 fractionalized publications threshold considered in the Leiden Ranking. While we identified some overlap between green and gold OA, the other three categories are exclusive from each other. Finally, in this paper we will consider as OA whichever document which adheres to any of these four types, however we will offer a deeper analysis to those following the green, gold and hybrid routes.

| Country | # univs | Total pubs | | Total OA | | Gold OA | | Green OA | | Hybrid OA | |
|---|---|---|---|---|---|---|---|---|---|---|---|
| | | Avg. | Std. Dev. | Avg. | Std. Dev. | Avg. | Std. Dev. | Avg. | Std. Dev. | Avg. | Std. Dev. |
| Netherlands | 13 | 13721,0 | 6209,6 | 7319,9 | 3645,1 = | 2005,8 | 1116,7 = | 6084,4 | 2974,0 ▼ | 1843,8 | 878,6 ▼ |
| United States | 173 | 10845,2 | 9676,8 | 5858,4 | 6060,4 ▼ | 1256,0 | 1216,3 ▼ | 4951,8 | 5329,3 ▼ | 1050,9 | 985,9 ▼ |
| Sweden | 11 | 10602,7 | 5747,9 | 5747,0 | 3425,7 ▼ | 1789,3 | 1186,8 ▲ | 4763,0 | 2768,4 ▼ | 1366,7 | 827,5 = |
| Australia | 26 | 10547,2 | 7854,3 | 4387,7 | 3615,7 ▼ | 1425,3 | 1139,3 = | 3474,8 | 2801,4 ▼ | 652,3 | 590,6 ▼ |
| Canada | 27 | 10418,7 | 9133,0 | 4342,5 | 4322,2 ▼ | 1378,1 | 1275,8 = | 3355,2 | 3400,5 ▼ | 692,6 | 716,3 ▼ |
| United Kingdom | 45 | 10116,0 | 8517,2 | 7178,2 | 6110,8 ▲ | 1620,5 | 1536,7 ▲ | 6546,4 | 5522,0 ▲ | 1957,7 | 1910,4 ▲ |
| France | 25 | 9518,0 | 6245,7 | 4871,0 | 3879,0 ▲ | 1281,8 | 941,3 ▲ | 4085,0 | 3439,5 ▲ | 755,9 | 577,2 ▲ |
| Germany | 50 | 7969,7 | 4689,9 | 3819,9 | 2499,5 = | 1268,0 | 866,6 ▲ | 3100,2 | 2057,5 = | 744,4 | 486,7 ▲ |
| China | 165 | 7249,0 | 7098,0 | 2153,2 | 2492,1 ▼ | 974,7 | 1044,5 ▼ | 1535,5 | 1998,8 ▼ | 405,2 | 503,0 ▼ |
| South Korea | 35 | 7029,5 | 5504,3 | 2444,5 | 2138,1 ▼ | 924,4 | 754,5 ▼ | 1827,3 | 1672,3 ▼ | 716,1 | 692,6 ▲ |
| Italy | 40 | 6881,4 | 4675,2 | 3005,8 | 2173,2 ▲ | 1043,7 | 739,1 = | 2515,5 | 1839,3 ▲ | 555,7 | 414,3 ▼ |
| Japan | 42 | 6322,2 | 6351,2 | 2841,2 | 3098,1 ▲ | 887,2 | 958,2 ▼ | 1960,1 | 2256,5 = | 567,6 | 608,5 ▲ |
| Brazil | 23 | 6248,0 | 6679,4 | 2346,2 | 2771,9 ▼ | 1198,1 | 1370,0 ▲ | 1703,4 | 2121,6 ▼ | 294,9 | 347,0 ▼ |
| Taiwan | 17 | 5783,7 | 3987,0 | 2130,2 | 1782,4 ▼ | 1095,6 | 880,8 ▲ | 1514,5 | 1326,0 ▼ | 465,4 | 423,5 ▲ |
| Spain | 34 | 5323,8 | 3395,0 | 2535,2 | 1807,6 ▲ | 731,3 | 571,6 = | 2220,6 | 1605,7 ▲ | 313,3 | 279,4 ▼ |
| Austria | 10 | 4702,3 | 2383,1 | 2481,9 | 1416,8 ▲ | 718,7 | 410,5 = | 2012,9 | 1193,2 ▲ | 747,2 | 349,5 ▲ |
| Iran | 26 | 3648,8 | 2244,3 | 546,5 | 429,2 ▼ | 219,5 | 181,7 ▼ | 343,9 | 295,2 ▼ | 88,8 | 87,3 ▼ |
| India | 25 | 3296,2 | 1492,7 | 632,2 | 438,0 ▼ | 236,8 | 175,9 ▼ | 461,2 | 373,3 ▼ | 107,7 | 83,1 ▼ |
| Poland | 24 | 3058,5 | 1646,7 | 1331,7 | 905,6 ▲ | 428,3 | 277,5 ▲ | 915,6 | 775,9 ▲ | 429,0 | 245,7 ▲ |
| Turkey | 20 | 2967,1 | 1271,7 | 885,9 | 461,5 ▲ | 291,3 | 161,4 ▲ | 584,4 | 429,8 ▲ | 140,1 | 89,0 ▲ |

**Figure 2. Institutional output at the country level for top 20 countries based on number of universities in Leiden Ranking for the 2014-2017 period. Countries are ordered based on average number of publications. Arrows show changes in ranking based on average of total number of publications**

**Results and discussion**

The two countries contributing most on the number of universities analysed are United States and China, more tripling the third and fourth countries (Figure 2). The Netherlands have on average the most productive universities followed by United States, Sweden and Australia. For the average number of OA publications, we observe that Netherlands is now followed by United Kingdom, United States and Sweden. British universities occupy higher positions when referring to total, green, gold or hybrid OA output on average. However, there seems to be large disparities within the country. On the other side we find countries such as Netherlands, Australia and Canada, which are in a lower position based on their average number of green and hybrid OA publications than what they should occupy, according to their overall average number of publications. We find differences on the size of the output of institutions by country.

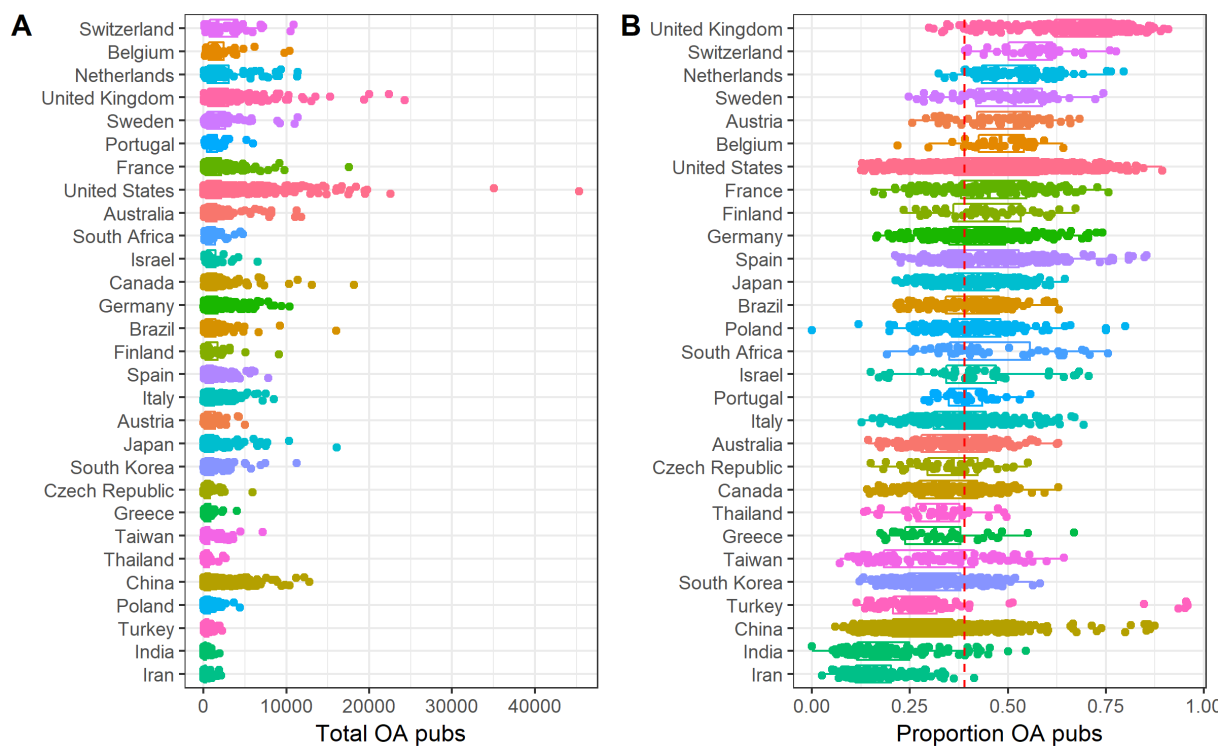

**Figure 3. Distribution of universities based on their A) total number and B) proportion of OA publications for countries with > 5 universities included. Countries ordered by median proportion of OA publications. Red dashed line shows world median.**

Disparities within countries are further analysed on figures 3 and 4. Figure 3 plots the distribution of universities by country for their total number and proportion of OA publications. Figure 4 does the same for green and gold OA. Overall, we observe that the United Kingdom is the country in which their universities are making a higher proportion of their publications openly available, followed by Switzerland and Sweden (Figure 3B). Furthermore, we find extreme cases in Turkey, China and Spain. In the two former cases, most of the universities show shares of OA lower than the world median. It is also worth noting that most of the countries with higher proportions of OA at the institutional level are European with two notable exceptions. These are South Africa and United States. In the case of the former, we find a very different pattern from the other two BRICS countries shown (China and India), which have a proportion of OA publications below world median. In the latter case, it does not occupy a leading position as it is usually the case with the United States. Russia is not present in these figures as only three universities from this country are included in the Leiden Ranking.

The ways in which universities are making their publications openly accessible varies greatly when distinguishing between gold and green OA. United Kingdom and Switzerland are again the ones with the highest median on the proportion of green OA their universities have (Figure 4B). It is also worth noting the great dispersion on proportion of green OA not only between countries but also within countries. While the world median proportion of green OA is 31.7%. It raises up to 67.4% for the United Kingdom and it is of 8.6% for Iran.

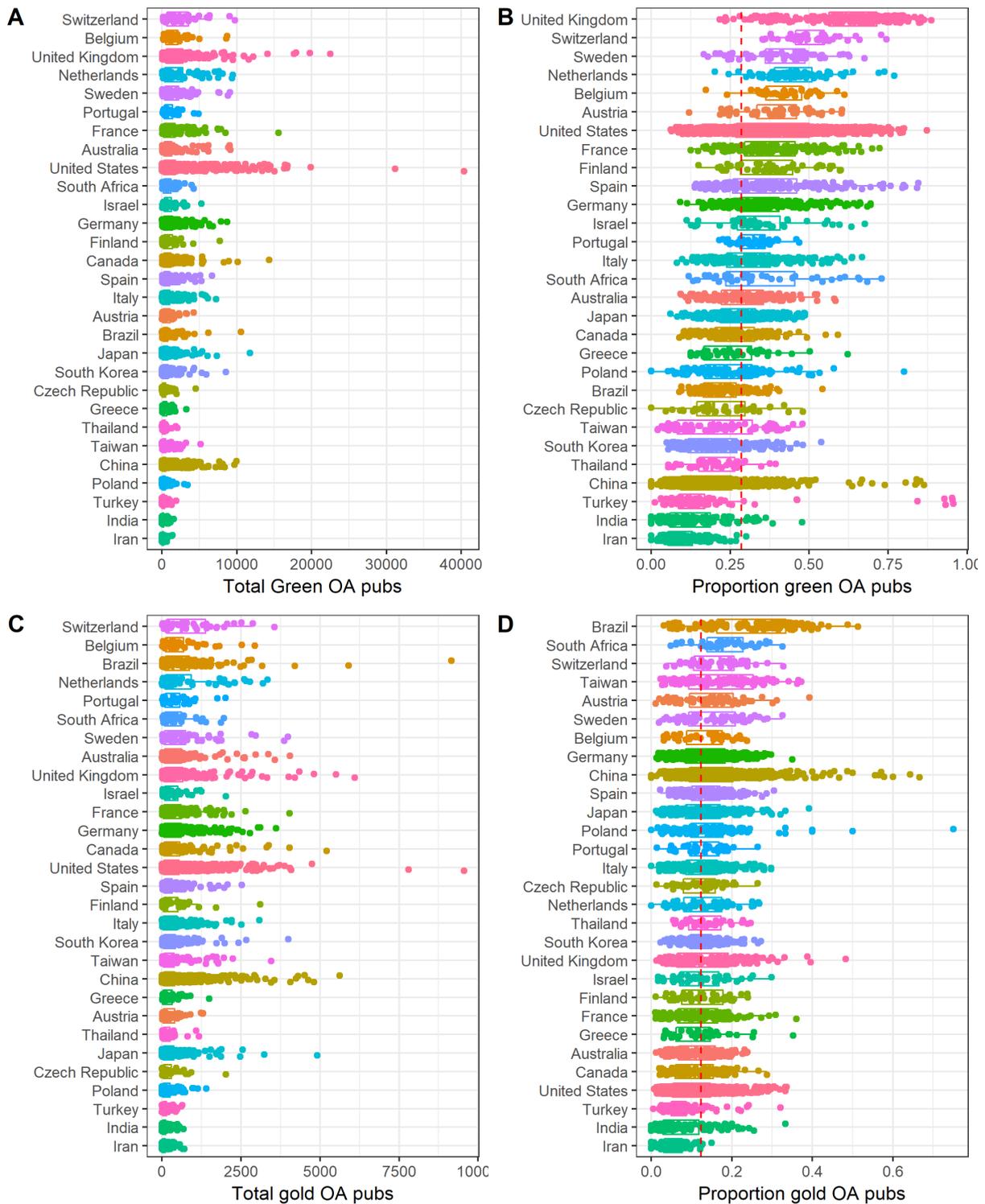

**Figure 4. Boxplots by country for countries > 5 universities included based on their A) total number and B) proportion of green OA pubs and C) total number and D) proportion of gold OA pubs. Countries ordered by median proportion of OA publications. Red dashed line shows world median.**

In the case of gold OA, the world median institutional share is 13.1%. As observed, there are less disparities within countries than in the case of green OA with the exception of China and Taiwan (figure 4D). In this case, Switzerland (19.0%), Brazil (18.3%) and South Africa (17.6%) are the countries with the largest proportion of their output in gold OA (median values). Taiwan is the country with greater disparities between its universities. To better interpret the patterns of these countries we look into the OA journal profile of these four countries, following the three models of gold OA proposed by Torres-Salinas, Robinson-Garcia, & Moed (2019). The first one refers to countries which publish in OA journals owned by publishing firms, preferably mega-journals and with a high Journal Impact Factor. The second model is that of countries which publish in OA journals edited in their own country, preferably in their native language and publicly funded. The third model is a mixed one where gold OA publications are channelled through both OA mega-journals and nationally oriented OA journals. In all cases PLoS One is the journal with the largest number of publications. In the case of Brazil, the rest of OA journals in the list are in a vast majority Brazilian journals listed in SCielo. In South Africa we observe a combination of regional journals and big OA publishers. Finally, Taiwan exhibits a greater reliance on journals from big OA publishers such as Nature Springer, PloS, MDPI or Hindawi.

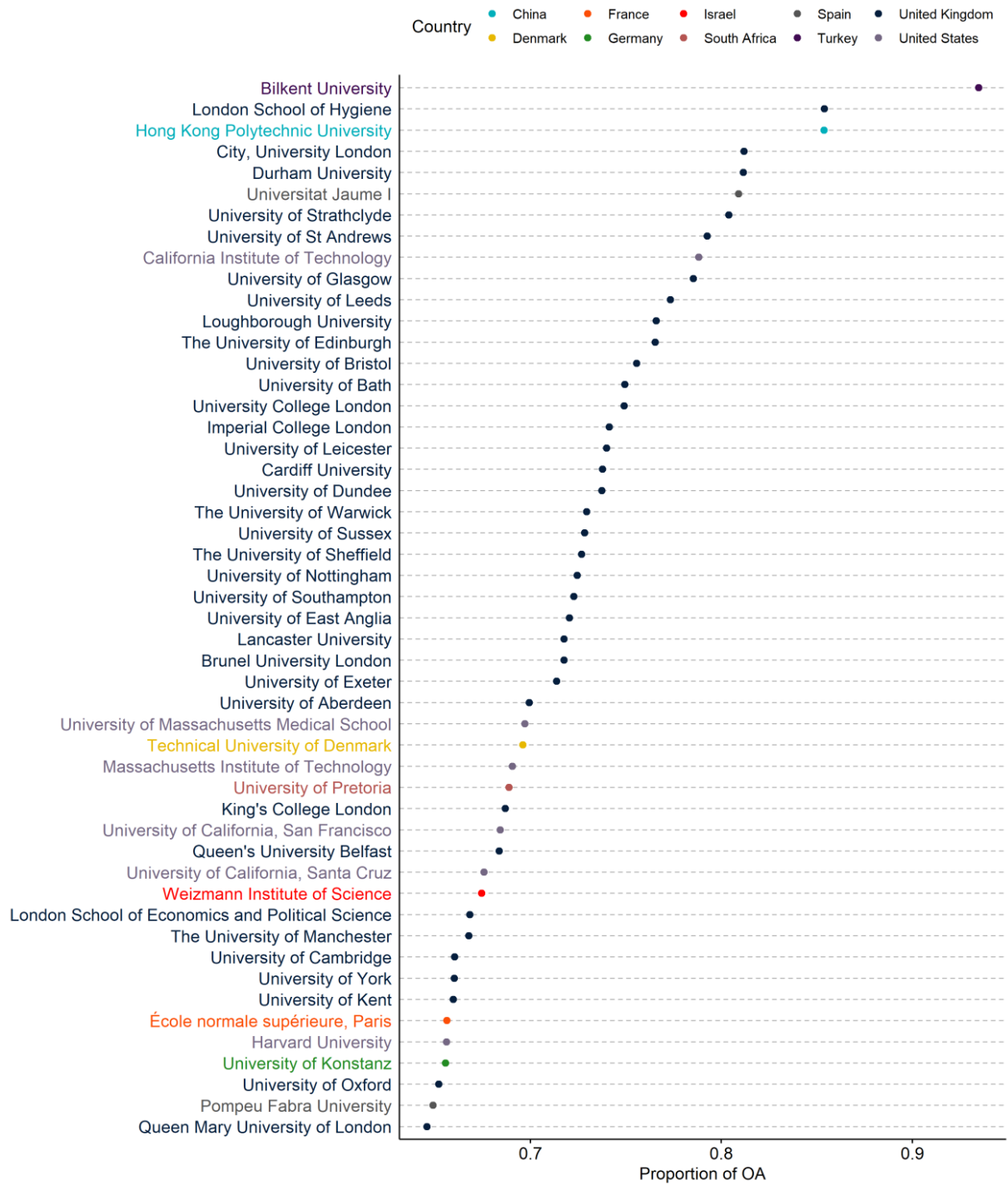

**Figure 5. Ranking of top 50 universities based on the proportion of Open Access publications in the 2014-2017 period**

Finally, we conclude by showcasing in figure 5 the top 50 universities with the largest proportion of OA worldwide. These 50 universities come from 10 different countries. More than half of them come from United Kingdom (35), including major universities such as London City University. Next, United States positions 6 universities, followed by Spain (2) and the remaining countries have one university each (Turkey, China, Denmark, South Africa, Germany, France and Israel) Here we note that the two outliers aforementioned from Turkey and China (figure 2B), are actually the top 2 universities on openly accessible literature, mostly relying on green OA (see figure 3B). In both cases, most of this literature is actually coming from their institutional repository.

**Concluding remarks**

This paper presents a first attempt at measuring OA uptake by universities worldwide. Europe is hardening its policies towards full OA, and initiatives such as Plan S are being supported by important international funding bodies (e.g., Wellcome Trust, Bill & Melinda Gates Foundation). The introduction of such policies may affect differently across Europe and such policies may expand to other countries. The inclusion of indicators on OA to the Leiden Ranking adds another dimension, which is less traditional, and focused on changes in scholarly communication practices. This can better inform how OA is being implemented and which routes are having a greater implementation. Finally, this contribution allows to study the distribution across the globe of OA uptake, to what extent the initial goals of the OA movement to distribute more equally over the globe reached, especially when looking at OA uptake in e.g., the Global South, effects of regulations (e.g., inclusion of hybrid OA by Plan S), etc.

Here we present a first attempt at developing OA indicators at the institutional level globally. However, there are many issues that still need to be dealt with. For instance, the consideration of Unpaywall as the most important means by which OA is captured, although welcome and remarkable, needs to be better assessed and understood (double occurrences, undetermined category, etc.). Also, it is important to understand better and make more distinct in OA analyses gold OA models and specifically publicly-funded gold OA (i.e., SCielo) versus APC models and private publishing firms.

**Acknowledgments**

Nicolas Robinson-Garcia is a Marie Sklodowska-Curie Experienced Researcher in the LEaDing Fellows COFUND programme from the European Commission. This research is partially funded by the South African DST-NRF Centre of Excellence in Scientometrics and Science, Technology and Innovation Policy (SciSTIP).